\documentclass[twocolumn]{aastex631}
\usepackage{newtxtext,newtxmath}
\usepackage[T1]{fontenc}

\usepackage{nicefrac}
\usepackage{amssymb}
\usepackage{latexsym}
\usepackage{amsmath, mathrsfs, bm}
\usepackage{eqnarray}
\usepackage{url}
\usepackage{cases}
\usepackage{epsfig}
\usepackage{color}
\usepackage{graphicx}
\usepackage{grffile}
\usepackage{float}
\usepackage{soul}
\usepackage{enumerate}
\usepackage{multirow}
\usepackage{threeparttable}
\usepackage{xspace}

\usepackage{bm}
\usepackage{times}

\usepackage{etoolbox}
\usepackage{multirow}

\begin{document}

\title{Constraints on Interacting Dark Energy Models from the DESI Baryon Acoustic Oscillation and DES Supernovae Data}

\author[0009-0004-6982-4021]{Tian-Nuo Li}
\affiliation{Key Laboratory of Cosmology and Astrophysics (Liaoning Province) \& Department of Physics, College of Sciences, Northeastern University, Shenyang 110819, People's Republic of China}

\author[0000-0002-0113-9499]{Peng-Ju Wu}
\affiliation{Key Laboratory of Cosmology and Astrophysics (Liaoning Province) \& Department of Physics, College of Sciences, Northeastern University, Shenyang 110819, People's Republic of China}

\author[0009-0005-6921-3201]{Guo-Hong Du}
\affiliation{Key Laboratory of Cosmology and Astrophysics (Liaoning Province) \& Department of Physics, College of Sciences, Northeastern University, Shenyang 110819, People's Republic of China}

\author[0000-0003-3697-3501]{Shang-Jie Jin}
\affiliation{Key Laboratory of Cosmology and Astrophysics (Liaoning Province) \& Department of Physics, College of Sciences, Northeastern University, Shenyang 110819, People's Republic of China}
\affiliation{Department of Physics, University of Western Australia, Perth WA 6009, Australia}

\author[0009-0008-8030-3461]{Hai-Li Li}
\affiliation{Basic Teaching Department, Shenyang Institute of Engineering, Shenyang 110136, People's Republic of China}

\author[0000-0002-3512-2804]{Jing-Fei Zhang}
\affiliation{Key Laboratory of Cosmology and Astrophysics (Liaoning Province) \& Department of Physics, College of Sciences, Northeastern University, Shenyang 110819, People's Republic of China}

\author[0000-0002-6029-1933]{Xin Zhang}
\affiliation{Key Laboratory of Cosmology and Astrophysics (Liaoning Province) \& Department of Physics, College of Sciences, Northeastern University, Shenyang 110819, People's Republic of China}
\affiliation{Key Laboratory of Data Analytics and Optimization for Smart Industry (Ministry of Education), Northeastern University, Shenyang 110819, People's Republic of China}
\affiliation{National Frontiers Science Center for Industrial Intelligence and Systems Optimization, Northeastern University, Shenyang 110819, People's Republic of China}

\correspondingauthor{Jing-Fei Zhang}
\email{jfzhang@mail.neu.edu.cn}

\correspondingauthor{Xin Zhang}
\email{zhangxin@mail.neu.edu.cn}

\begin{abstract}
The recent results from the first-year baryon acoustic oscillations (BAO) data released by the Dark Energy Spectroscopic Instrument (DESI), combined with cosmic microwave background (CMB) and Type Ia supernova (SN) data, have shown a detection of significant deviation from a cosmological constant for dark energy. In this work, we utilize the latest DESI BAO data in combination with the SN data from the full 5 yr observations of the Dark Energy Survey and the CMB data from the Planck satellite to explore potential interactions between dark energy and dark matter. We consider four typical forms of the interaction term $Q$. Our findings suggest that interacting dark energy (IDE) models with $Q \propto \rho_{\rm de}$ support the presence of an interaction where dark energy decays into dark matter. Specifically, the deviation from $\Lambda$CDM for the IDE model with $Q=\beta H_0\rho_{\rm de}$ reaches the $3\sigma$ level. These models yield a lower value of Akaike information criterion than the $\Lambda$CDM model, indicating a preference for these IDE models based on the current observational data. For IDE models with $Q\propto\rho_{\rm c}$, the existence of interaction depends on the form of the proportionality coefficient $\Gamma$. The IDE model with $Q=\beta H\rho_{\rm c}$ yields $\beta=0.0003\pm 0.0011$, which essentially does not support the presence of the interaction. In general, whether the observational data support the existence of interaction is closely related to the model. Our analysis helps to elucidate which type of IDE model can better explain the current observational data.

\end{abstract}


\section{Introduction}
In 1998, two independent studies of distant Type Ia supernovae (SNe) discovered the accelerated expansion of the Universe \citep{SupernovaSearchTeam:1998fmf,SupernovaCosmologyProject:1998vns}, which was further independently confirmed by observations of the cosmic microwave background (CMB; \citealp{WMAP:2003ivt}; \citealp{WMAP:2003elm}; \citealp{Planck:2018vyg}) and baryon acoustic oscillations (BAO; \citealp{SDSS:2005xqv,BOSS:2016wmc,eBOSS:2020yzd}). In order to explain the cosmic
acceleration, the concept of ``dark energy,'' which is an exotic form of energy with negative
pressure, has been proposed. At present, dark energy occupies about 68\% of the total energy density of the cosmos, dominating the evolution of the current Universe. The cosmological constant $\Lambda$, proposed by Einstein in 1917, has been regarded as the most straightforward candidate for dark energy up to now. For a long time, the $\Lambda$CDM model, which is primarily composed of a cosmological constant $\Lambda$ and cold dark matter (CDM), has been considered a standard model of cosmology.

The $\Lambda$CDM model fits the CMB data quite well, and the six basic parameters have been constrained with unprecedented precision. However, as the measurement precision of the cosmological parameters improves, some puzzling issues have appeared. For the measurement of the Hubble constant, especially, the local distance ladder method gives $H_0=73.04\pm1.04$ ${\rm km}~{\rm s}^{-1}~{\rm Mpc}^{-1}$ \citep{Riess:2021jrx}, while the CMB observation infers $H_0=67.36\pm0.54$ ${\rm km}~{\rm s}^{-1}~{\rm Mpc}^{-1}$ \citep{Planck:2018vyg}, revealing a tension above $5\sigma$. Recently, the Hubble tension has been widely discussed in the literature \citep{Li:2013dha,Zhang:2014dxk,Feng:2017nss,Zhao:2017urm,Verde:2019ivm,Riess:2019qba,Guo:2018ans,Guo:2019dui,Gao:2021xnk,DiValentino:2021izs,Cai:2021wgv,Kamionkowski:2022pkx,Gao:2022ahg,Lynch:2024hzh} (for some relevant forecast analyses, see also, e.g.,~\citealp{Zhao:2010sz,Cai:2017aea,Du:2018tia,Zhang:2019ylr,Zhang:2019loq,Chen:2020dyt,Chen:2020zoq,Bian:2021ini,Jin:2021pcv,Wang:2021srv,Wang:2022oou,Zhao:2022yiv,Jin:2022qnj,Jin:2023zhi,Zhang:2023gye,Dong:2024bvw,Jin:2023sfc,Jin:2023tou,Song:2022siz,Xiao:2024nmi}). On the other hand, the cosmological constant $\Lambda$ in the $\Lambda$CDM model, which is equivalent to the vacuum energy density, has also been suffering from serious theoretical challenges, namely, the ``fine-tuning'' and ``cosmic coincidence'' problems \citep{Sahni:1999gb,Bean:2005ru}. Therefore, it is far-fetched to take the $\Lambda$CDM model as the eventual cosmological scenario. All of these facts imply that we need to consider new physics beyond the $\Lambda$CDM model. 

Among the different extensions to the $\Lambda$CDM model, there exists a category known as interacting dark energy (IDE) models, in which the direct and nongravitational interaction between dark energy and dark matter is considered. The IDE models have been extensively discussed in the literature \citep{Zhang:2005rj,Zhang:2007uh,Li:2009zs,Zhang:2009qa,Li:2010ak,Li:2011ga,Fu:2011ab,Zhang:2012uu,Cui:2015ueu,Szydlowski:2008by,Zhang:2017ize,Li:2019ajo,DiValentino:2019ffd,Feng:2019jqa,Zhang:2021yof,Wang:2021kxc,Jin:2022tdf,Nunes:2022bhn,Zhao:2022bpd,Han:2023exn,Forconi:2023hsj,Li:2023gtu,Halder:2024uao,Benisty:2024lmj,Nong:2024bkr}. IDE models can not only help alleviate the Hubble tension \citep{Yang:2018uae,Guo:2018ans,Vagnozzi:2019ezj} and the coincidence problem of dark energy \citep{Comelli:2003cv,Cai:2004dk,Zhang:2005rg} but also help probe the fundamental nature of dark energy and dark matter. In fact, the CMB data alone can only measure the six basic parameters at high precision for the $\Lambda$CDM model, and when the model is extended to include new parameters, the CMB data alone cannot provide precise measurements for them. This is because the newly extended parameters beyond the $\Lambda$CDM model will severely degenerate with other parameters. Hence, low-redshift cosmological probes such as the BAO and SN observations need to be combined with the CMB data to break the cosmological parameter degeneracies. 

Recently, the Dark Energy Spectroscopic Instrument (DESI) collaboration released the first-year data from the measurements of BAO in galaxy, quasar, and Ly$\alpha$ forest tracers \citep{DESI:2024uvr,DESI:2024lzq}, and the cosmological parameter results derived from these measurements \citep{DESI:2024mwx}. For the cosmological constraints, the DESI collaboration provided cosmological inference results for various cosmological models, including the $\Lambda$CDM model, the $w$CDM model, and the $w_0w_a$CDM model. In particular, for the $w_0w_a$CDM model, combining the DESI BAO data with the CMB and DESY5 data leads to deviations from $\Lambda$CDM at up to $3.9\sigma$. Interestingly, \citet{Park:2024jns} have also reported similar evidence for dynamical dark energy using the $w_0w_a$CDM model with other BAO data. The deviations from the $\Lambda$CDM model observed in the DESI BAO data have sparked extensive discussions regarding dark energy. Recently, several studies have sought to constrain various aspects of cosmological physics using the DESI BAO data (see, e.g., \citealp{Wang:2024rjd,Wang:2024hks,Giare:2024syw,Giare:2024gpk,Gomez-Valent:2024tdb,Escamilla-Rivera:2024sae,DiValentino:2024xsv,Colgain:2024xqj,Qu:2024lpx,Du:2024pai,Wang:2024dka,Allali:2024anb,Reboucas:2024smm,Jiang:2024viw,Escamilla:2024ahl,Sabogal:2024yha,Cortes:2024lgw,DESI:2024aqx,Toda:2024ncp,Yang:2024kdo,Pang:2024qyh,Wang:2024pui}). 

Additionally, the DESI BAO data, along with other data like CMB and SN have been considered to constrain the IDE model \citep{Giare:2024smz}. However, in the study of IDE models, the choice of different forms of interaction terms affects the results. Hence, in this work, we use the current observational data including the DESI BAO data, the CMB data from Planck, and the SN data from the Dark Energy Survey (DES), to constrain IDE models by considering four typical forms of interaction term $Q$. Our motivation is to explore whether there is an interaction between dark energy and dark matter and which types of IDE models are more supported by the current cosmological observations.

This work is organized as follows: In Section.~\ref{sec2}, we briefly introduce the IDE models and cosmological data used in this work. In Section.~\ref{sec3}, we report the constraint results and make some relevant discussions. The conclusion is given in Section.~\ref{sec4}.

\section{ models and data}\label{sec2}
\subsection{Interacting dark energy models}

In a flat Friedmann--Roberston--Walker universe, the Friedmann equation is given by
\begin{equation}\label{2.1}
3M^2_{\rm{pl}} H^2=\rho_{\rm{de}}+\rho_{\rm c}+\rho_{\rm b}+\rho_{\rm r},
\end{equation}
where $3M^2_{\rm{pl}} H^2$ is the critical density of the Universe, and $\rho_{\rm{de}}$, $\rho_{\rm c}$, $\rho_{\rm b}$, and $\rho_{\rm r}$ represent the energy densities of dark energy, CDM, baryon, and radiation, respectively.
In the IDE models, if assuming a direct interaction between dark energy and CDM, the energy conservation equations can be given by
\begin{align}\label{conservation1}
\dot{\rho}_{\rm de} +3H(1+w)\rho_{\rm de}= Q,\\
\dot{\rho}_{\rm c} +3 H \rho_{\rm c}= -Q,
\end{align}
where the dot is the derivative with respect to the cosmic time $t$, $H$ is the Hubble parameter, $w$ is the equation of state of dark energy, and $Q$ denotes the phenomenological interaction term describing the energy transfer rate between dark
energy and dark matter due to the interaction. For the phenomenological IDE models, the form of $Q$ is usually assumed to be proportional to the energy density of dark energy or CDM \citep{Amendola:1999qq,Billyard:2000bh}. To balance the dimensions, it must be multiplied by a quantity with units of the inverse of time. The obvious choice is the Hubble parameter $H$, as it can provide an analytical solution for the conservation equations. Thus, the interaction between dark energy and dark matter can be expressed phenomenologically in forms such as $Q=\beta H\rho_{\rm c}$ or $Q=\beta H\rho_{\rm de}$. However, in the research area of IDE, there is another perspective that $Q$ should not involve the Hubble parameter $H$ because the local interaction should not depend on the global expansion of the Universe \citep{Boehmer:2008av,Valiviita:2008iv,Caldera-Cabral:2008yyo,He:2008si,Clemson:2011an}. Thus, according to this perspective, another form of $Q$ is assumed, such as $Q=\beta H_0\rho_{\rm c}$ or $Q=\beta H_0\rho_{\rm de}$, where the appearance of the $H_0$ is only for a dimensional considerations. The proportionality coefficient $\Gamma$ has the dimension of energy, and thus takes the form of $\Gamma=\beta H$ or $\Gamma=\beta H_0$, where $\beta$ is the dimensionless coupling parameter. $\beta > 0$ indicates CDM decaying into dark energy, $\beta < 0$ indicates dark energy decaying into CDM, and $\beta = 0$ indicates no interaction between dark energy and CDM. In this work, we do not aim to study the nature of dark energy purely from a theoretical perspective but instead approach the issue of dark energy phenomenologically. We focus solely on the case where $w=-1$ to avoid introducing additional parameters, and the corresponding IDE model is denoted as the I$\Lambda$CDM model. Hence, we consider four typical phenomenological forms of $Q$: $Q=\beta H\rho_{\rm de}$ (I$\Lambda$CDM1), $Q=\beta H\rho_{\rm c}$ (I$\Lambda$CDM2), $Q=\beta H_0\rho_{\rm de}$ (I$\Lambda$CDM3), and $Q=\beta H_0\rho_{\rm c}$ (I$\Lambda$CDM4), in I$\Lambda$CDM models. 

In recent years, it has been found that IDE models may experience an early-time large-scale instability problem \citep{Majerotto:2009zz,Clemson:2011an}. This instability arises because cosmological perturbations of dark energy within IDE models diverge in certain regions of parameter space, potentially leading to the breakdown of IDE cosmology at the perturbation level. To avoid this problem, the parameterized post-Friedmann (PPF) approach \citep{Fang:2008sn,Hu:2008zd} was extended to the IDE models \citep{Li:2014eha,Li:2014cee,Li:2015vla,Li:2023fdk}, referred to as the ePPF approach. This ePPF approach can safely calculate the cosmological perturbations in the whole parameter space of the IDE models. In this work, we employ the ePPF approach to treat the cosmological perturbations (see, e.g.,~\citealp{Zhang:2017ize,Feng:2018yew}, for applications of the ePPF approach).
\subsection{Cosmological data}
\begin{table*}[!htb]
\setlength\tabcolsep{14pt}
\renewcommand{\arraystretch}{1.5}
\centering
\caption{\label{tab1} Statistics of the DESI Samples Utilized in the DESI DR1 BAO Measurements for This Paper.}
\begin{tabular}{cccccc}
\hline
\hline Tracer &Redshift & $z_{\rm eff}$ & $D_{\mathrm{M}}/r_{\mathrm{d}}$ & $D_{\mathrm{H}}/r_{\mathrm{d}}$ & $D_{\mathrm{V}}/r_{\mathrm{d}}$   \\
\hline

BGS   & $0.1-0.4$ & $0.30$  &...&... & $7.93 \pm 0.15$  \\
LRG1  & $0.4-0.6$ & $0.51$  & $13.62 \pm 0.25$ & $20.98 \pm 0.61$ &...      \\
LRG2  & $0.6-0.8$ & $0.71$  & $16.85 \pm 0.32$ & $20.08 \pm 0.60$ &...        \\
LRG3+ELG1  & $0.8-1.1$ & $0.93$  & $21.71 \pm 0.28$ & $17.88 \pm 0.35$ &...   \\
ELG2  & $1.1-1.6$ & $1.32$  & $27.79 \pm 0.69$ & $13.82 \pm 0.42$ &...     \\
QSO   & $0.8-2.1$ & $1.49$   &...&... & $26.07 \pm 0.67$             \\
Lya QSO  & $1.77-4.16$ & $2.33$   & $39.71 \pm 0.94$ & $8.52 \pm 0.17$ &...   \\ 
\hline
\end{tabular}
\end{table*}

In this work, we employ the Markov Chain Monte Carlo (MCMC) package {\tt CosmoMC} to infer the posterior distributions of parameters \citep{Lewis:2002ah,Lewis:2013hha}. We assess the convergence of the MCMC chains using the Gelman-Rubin statistics quantity $R - 1 < 0.02$ \citep{Gelman:1992zz}. The MCMC chains are analyzed using the public package {\tt Getdist} \citep{Lewis:2019xzd}. The basic parameter space of the IDE models is \{$\Omega_b h^2$, $\Omega_c h^2$, $\log(10^{10} A_{\mathrm{s}})$, $100\theta_\mathrm{MC}$, $n_{\mathrm{s}}$, $\tau_{\rm reio}$, $\beta$\}. We use the current observational data to constrain the IDE models and obtain the best-fit values and the $1\sigma$--$2\sigma$ confidence level ranges for the parameters of interest \{$H_{0}$, $\Omega_{\mathrm{m}}$, $\beta$\}. We adopt the following observational data sets:

$1.$ \text{\textit{Cosmic Microwave Background.}} The Planck satellite has provided precise measurements of the power spectra of anisotropies in the CMB \citep{Planck:2018nkj,Planck:2018vyg}. These observations hold significant importance for cosmology as they have revealed details about the matter composition, topology, and large-scale structure of the Universe. We adopt the Planck temperature (TT) and polarization (EE) auto-spectra, as well as their cross-spectra (TE) at $\ell\geq 30$, the low-$\ell$ Commander temperature likelihood, and the low-$\ell$ SimAll EE likelihood from the Planck 2018 data release \citep{Planck:2018vyg}.\footnote{The likelihood is available at \url{https://pla.esac.esa.int}.} We also conservatively employ
the Planck Public Release 4 (PR4) lensing likelihood from a combination of NPIPE PR4 Planck CMB lensing reconstruction \citep{Carron:2022eyg}.\footnote{The likelihood is available at \url{https://github.com/carronj/planck_PR4_lensing}.} We label the Planck 2018 dataset and the Planck PR4 lensing dataset as CMB.

$2.$ \text{\textit{Baryon Acoustic Oscillations.}} BAO provides a standard ruler for a typical length scale to measure the angular diameter
distance $D_{\mathrm{A}}(z)$ and Hubble parameter $H(z)$. BAO can be measured by analyzing the spatial distribution of
galaxies and consequently provides an independent way to measure the expansion rate of the Universe and can impose significant constraints on the cosmological parameters \citep{Weinberg:2013agg,eBOSS:2020yzd,Wu:2021vfz,Wu:2022jkf}. The DESI BAO data include tracers of the bright galaxy sample (BGS), luminous red galaxies (LRG), emission line galaxies (ELG), quasars (QSO), and the Ly$\alpha$ forest in a redshift range $0.1\leq z \leq 4.2$ \citep{DESI:2024uvr,DESI:2024lzq}. These tracers are described through the transverse comoving distance $D_{\mathrm{M}}/r_{\mathrm{d}}$, the angle-averaged distance $D_{\mathrm{V}}/r_{\mathrm{d}}$, and the Hubble horizon $D_{\mathrm{H}}/r_{\mathrm{d}}$, where  $r_{\mathrm{d}}$ is the comoving sound horizon at the drag epoch. We use 12 DESI BAO measurements in Table~\ref{tab1} based on~\cite{DESI:2024mwx}.\footnote{The DESI BAO data used in this work was made public with Data Release 1 (details at \url{https://data.desi.lbl.gov/doc/releases/}).} We label the DESI BAO data  as DESI.

$3.$ \text{\textit{Type Ia Supernovae.}} SNe provide measurements of luminosity distances $D_{\mathrm{L}}(z) = (1+z) D_{\mathrm{M}}(z)$. The luminosity distances are powerful distance indicators to probe the background evolution of the Universe, especially the equation of state of dark energy. Recently, the DES collaboration released part of the full 5 yr data based on a new, homogeneously selected sample of 1635 photometrically classified SNe (with redshifts in the range $0.1 < z < 1.3$), complemented by 194 low-redshift SNe (with redshifts in the range $0.025 < z < 0.1$), totaling 1829 SNe. This sample quintuples the number of high-quality $z > 0.5$ SNe compared to the previous leading Pantheon+ compilation and results in the tightest cosmological constraints achieved by any SN data set to date \citep{DES:2024jxu}. We label the SN data  as DESY5.\footnote{Data available at \url{https://github.com/des-science/DES-SN5YR}.}

\begin{table*}[!htb]
\renewcommand\arraystretch{1.5}
\centering
\caption{Fitting Results (68.3\% Confidence Level) in the $\Lambda$CDM and IDE Models from the CMB, CMB+DESI, and CMB+DESI+DESY5 Data.}
\label{tab2}
\begin{tabular}{lcccccc} 
\hline
\hline
Data&Parameter & $\Lambda$CDM & I$\Lambda$CDM1 & I$\Lambda$CDM2 & I$\Lambda$CDM3 & I$\Lambda$CDM4\\ 
\hline
CMB	& $H_{0}$ & $67.28\pm 0.52$ & $66.12^{+5.10}_{-4.20}$ & $65.90\pm 1.51$ & $66.21^{+3.50}_{-3.10}$ & $ 64.45\pm 3.20$\\
  & $\Omega_{\mathrm{m}}$ & $0.316\pm 0.007$ & $0.350^{+0.130}_{-0.161}$ & $0.333^{+0.020}_{-0.022}$ & $0.351^{+0.122}_{-0.150}$ & $0.388^{+0.055}_{-0.091}$\\
& $\beta$ & $...$ & $-0.1310^{+0.5500}_{-0.3600}$ & $ -0.0018\pm 0.0022$ & $-0.1530^{+0.7400}_{-0.3900}$ & $ -0.1100^{+0.1500}_{-0.1300}$\\

CMB+DESI & $H_{0}$ & $68.05\pm 0.41$ & $70.50^{+1.51}_{-1.80}$ & $68.72\pm 0.60$ & $ 70.4\pm 1.82$ & $ 69.4\pm 1.04$\\
& $\Omega_{\mathrm{m}}$ & $0.305\pm 0.005$ & $0.219^{+0.061}_{-0.050}$ & $ 0.298\pm 0.007$ & $ 0.208\pm 0.075$& $0.279^{+0.017}_{-0.020}$\\
& $\beta$ & $...$ & $ 0.2900\pm 0.1810$ & $ 0.0011\pm 0.0013$ & $ 0.3600^{+0.3700}_{-0.2800}$ & $ 0.0750\pm 0.0520$\\

CMB+DESI+DESY5 & $H_{0}$ & $67.60\pm 0.37$ & $ 66.18\pm 0.63 $ & $ 67.72\pm 0.56$ & $ 65.94\pm 0.63$ & $ 66.59\pm 0.69$\\
& $\Omega_{\mathrm{m}}$ & $0.311\pm 0.005$ & $0.367\pm 0.020$ & $0.310\pm 0.007$ & $0.393\pm 0.026$ & $0.335\pm 0.015$\\
& $\beta$ & $...$ & $ -0.1970\pm 0.0710$ & $ 0.0003\pm 0.0011$ & $ -0.4000\pm 0.1300$ & $ -0.0670\pm 0.0380$\\ 
&$\chi_{\rm min}^2$&$4493.706$ & $4484.452$ & $4492.344$ & $4481.438$ & $4489.528$\\
&$\Delta$ AIC& $0$ & $-7.254$ & $0.638$ & $-10.628$ & $-2.718$\\
  
\hline
\end{tabular}
\begin{threeparttable}
\begin{tablenotes}[flushleft]
      \item \textbf{Note.} Here, $H_{0}$ is expressed in units of ${\rm km}~{\rm s}^{-1}~{\rm Mpc}^{-1}$.
    \end{tablenotes}
\end{threeparttable}
\end{table*}

\section{Results and discussions}\label{sec3}

\begin{figure*}[!htp]
\includegraphics[width=0.45\textwidth]{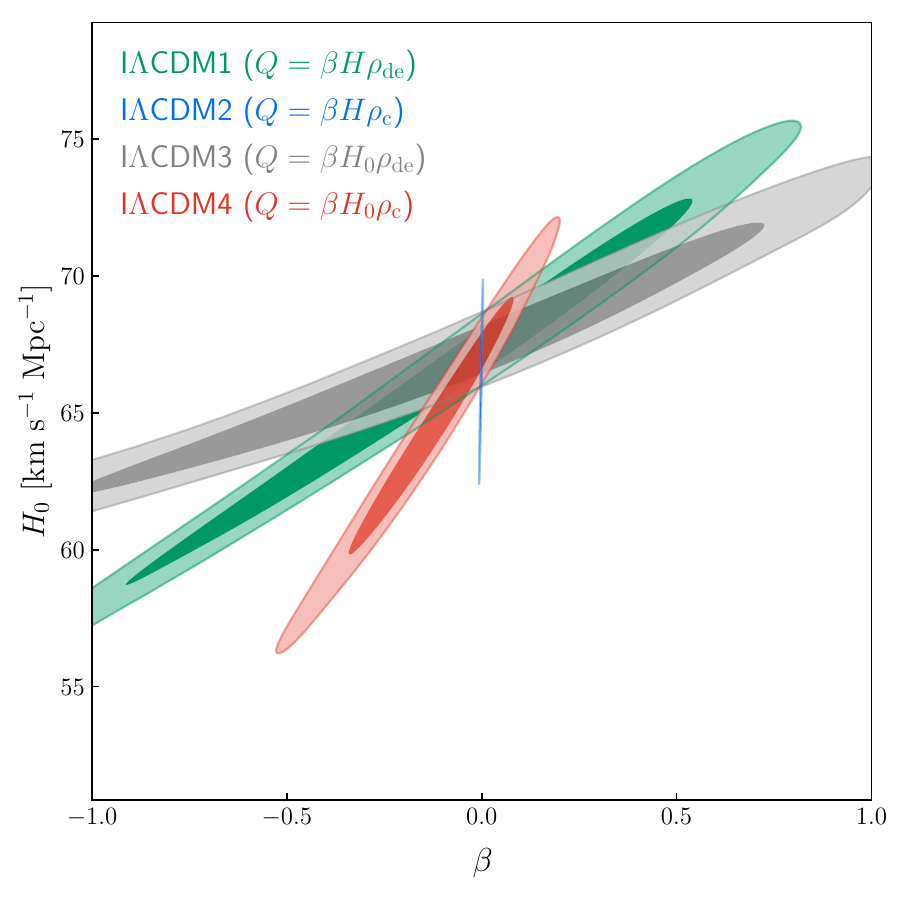} \ \hspace{1cm}
\includegraphics[width=0.45\textwidth]{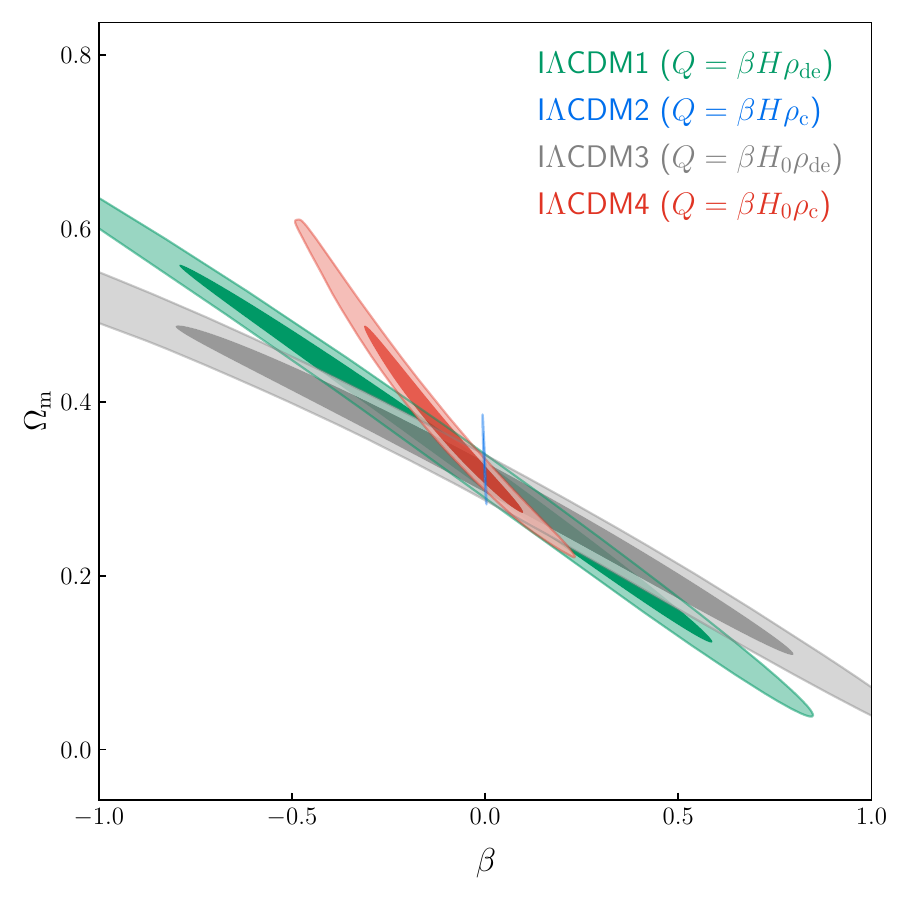}\ \hspace{1cm}
\centering \caption{\label{fig1} Two-dimensional marginalized contours (68.3\% and 95.4\% confidence level) in the $\beta$--$H_0$ and $\beta$--$\Omega_{\rm m}$ planes by using the CMB data in the I$\Lambda$CDM1, I$\Lambda$CDM2, I$\Lambda$CDM3, and I$\Lambda$CDM4 models.}
\end{figure*}
\begin{figure*}[!htp]
\includegraphics[width=0.45\textwidth]{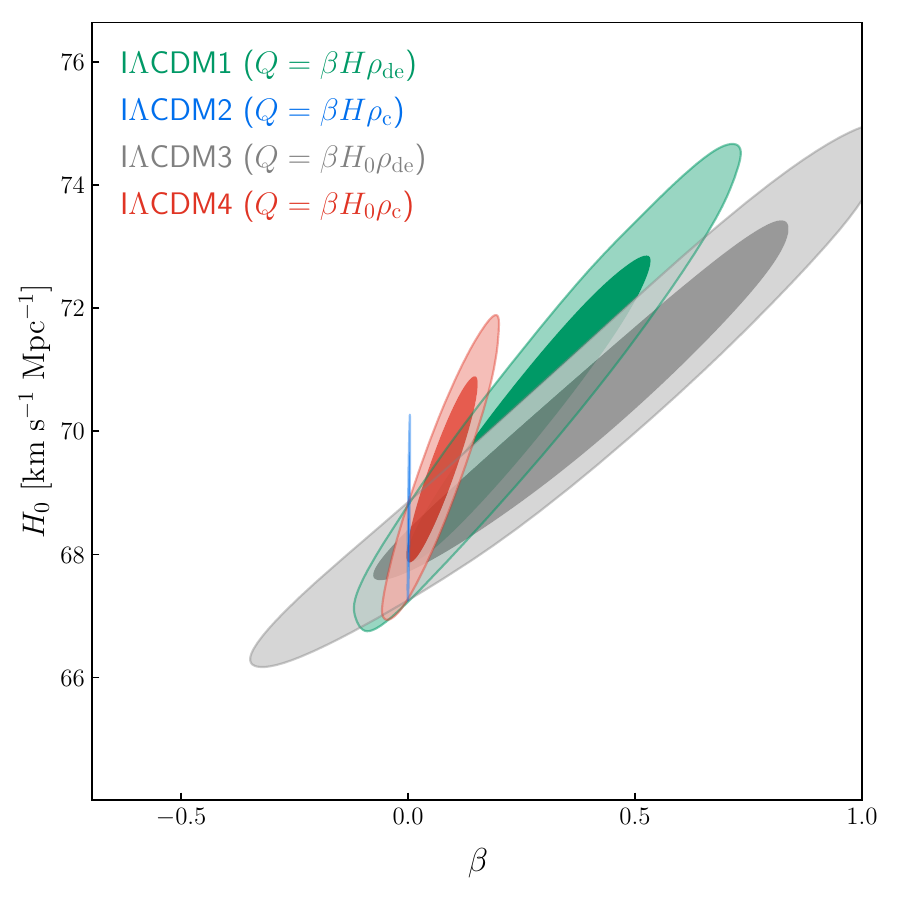} \ \hspace{1cm}
\includegraphics[width=0.45\textwidth]{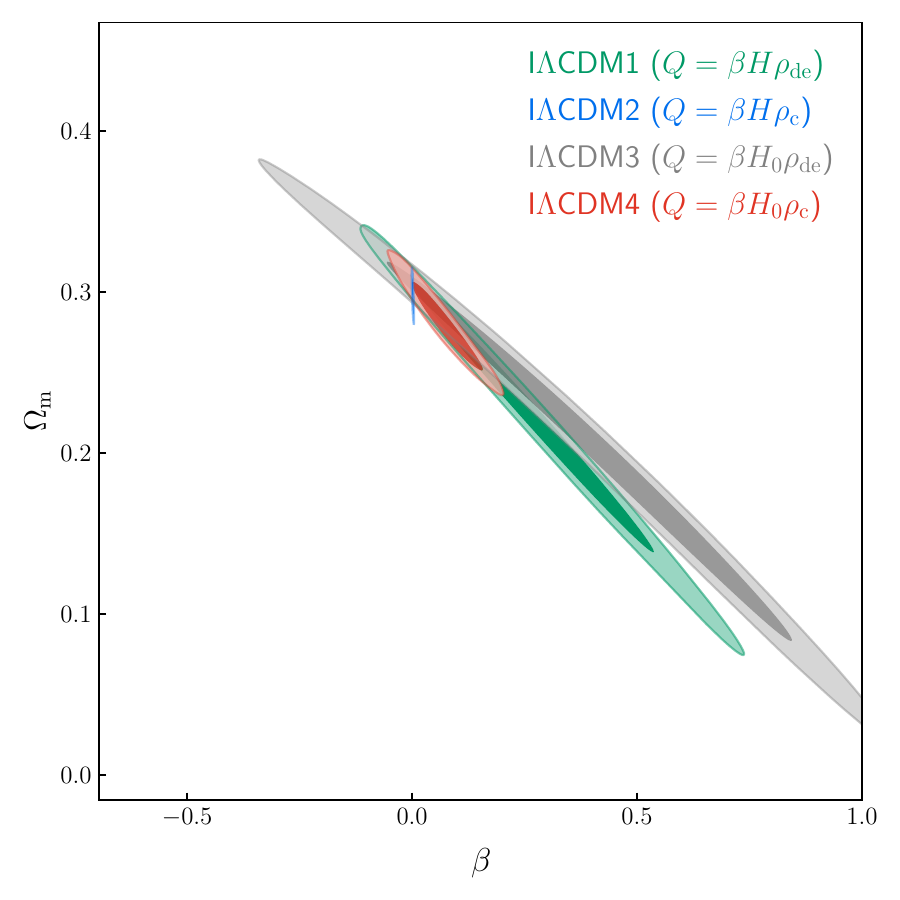}\ \hspace{1cm}
\centering \caption{\label{fig2} Two-dimensional marginalized contours (68.3\% and 95.4\% confidence level) in the $\beta$--$H_0$ and $\beta$--$\Omega_{\rm m}$ planes by using the CMB+DESI data in the I$\Lambda$CDM1, I$\Lambda$CDM2, I$\Lambda$CDM3, and I$\Lambda$CDM4 models.}
\end{figure*}
\begin{figure*}[!htp]
\includegraphics[width=0.45\textwidth]{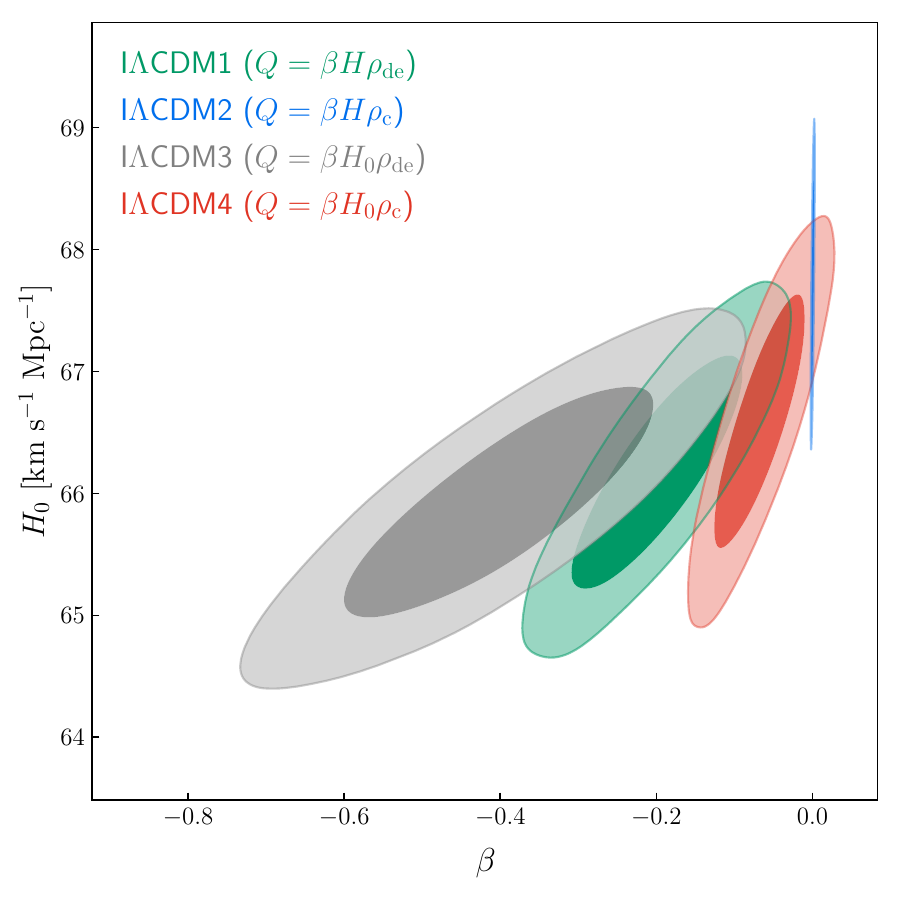} \ \hspace{1cm}
\includegraphics[width=0.45\textwidth]{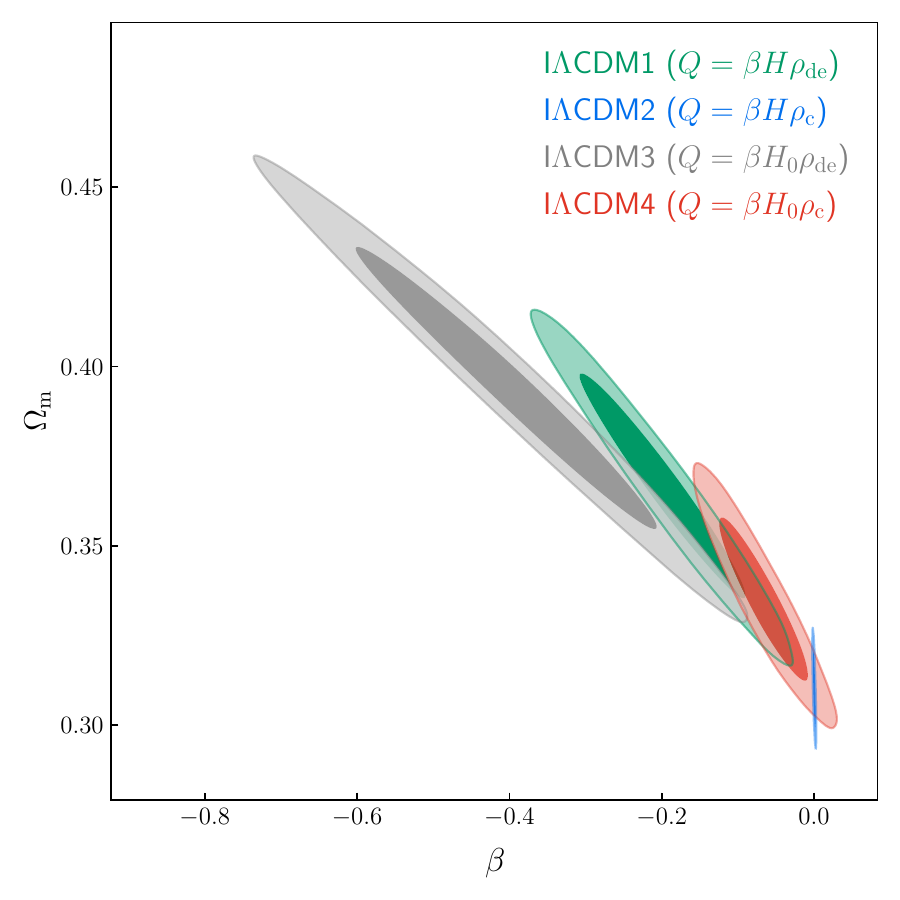}\ \hspace{1cm}
\centering \caption{\label{fig3} Two-dimensional marginalized contours (68.3\% and 95.4\% confidence level) in the $\beta$--$H_0$ and $\beta$--$\Omega_{\rm m}$ planes by using the CMB+DESI+DESY5 data in the I$\Lambda$CDM1, I$\Lambda$CDM2, I$\Lambda$CDM3, and I$\Lambda$CDM4 models.}
\end{figure*}
\begin{figure*}[htbp]
\includegraphics[scale=0.4]{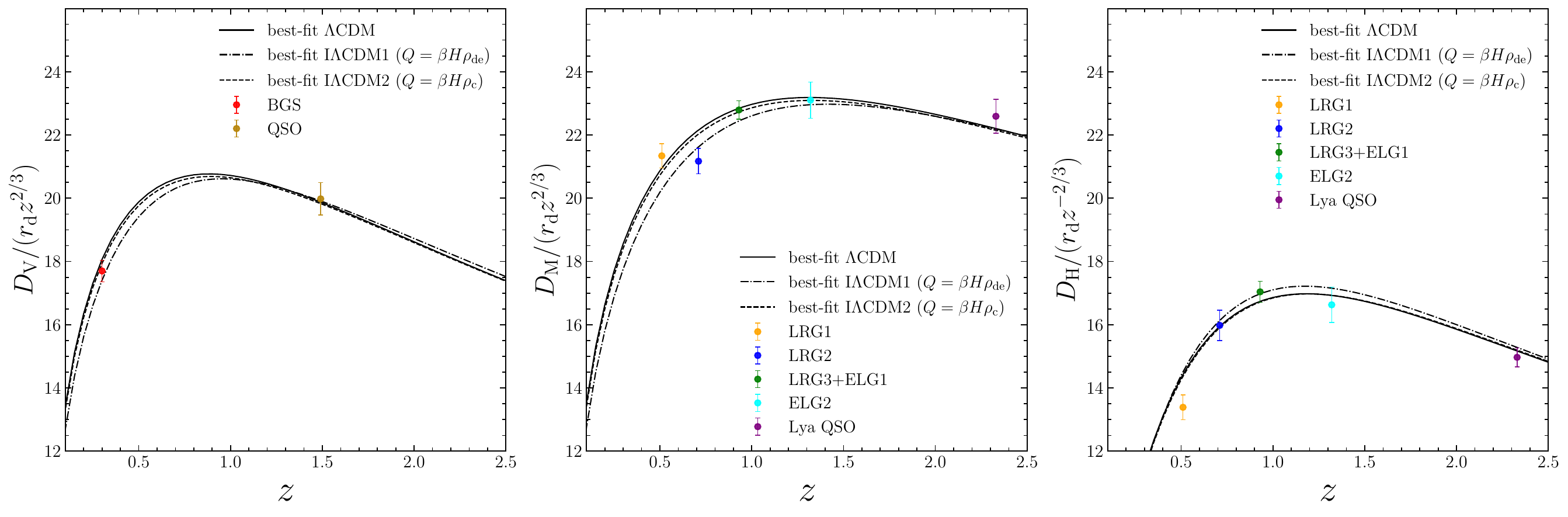}
\centering
\caption{\label{fig4} Best-fit predictions for distance-redshift relations for $\Lambda$CDM (solid line), I$\Lambda$CDM1 (dotted-dashed line), and I$\Lambda$CDM2 (dashed line) obtained from the analysis of CMB+DESI data. For visual clarity and to compress the dynamic range of the plot, we applied an arbitrary scaling. The error bars represent $\pm 1\sigma$ uncertainties.}
\end{figure*}

In this section, we shall report the constraint results of the cosmological parameters. We consider the I$\Lambda$CDM1 ($Q=\beta H\rho_{\rm de}$), I$\Lambda$CDM2 ($Q=\beta H\rho_{\rm c}$), I$\Lambda$CDM3 ($Q=\beta H_0\rho_{\rm de}$), and I$\Lambda$CDM4 ($Q=\beta H_0\rho_{\rm c}$) models to perform a cosmological analysis using current observational data, including DESI, CMB, and DESY5 data. We show the $1\sigma$ and $2\sigma$ posterior distribution contours for various cosmological parameters in the four IDE models, as shown in Figures~\ref{fig1}--\ref{fig3}. The $1\sigma$ errors for the marginalized parameter constraints are summarized in Table~\ref{tab2}. We compare the best-fit predictions for the three different types of (rescaled) distances obtained by DESI BAO measurements using CMB+DESI data in the $\Lambda$CDM, I$\Lambda$CDM1, and I$\Lambda$CDM2 models, as shown in Figure~\ref{fig4}.

In Figure~\ref{fig1}, we show the constraint results of the CMB data in the $\beta$--$H_0$ (left panel) and $\beta$--$\Omega_{\rm m}$ (right panel) planes for the four IDE models. When using the CMB data alone, the constraint values of $\beta$ are $-0.1310^{+0.5500}_{-0.3600}$ (I$\Lambda$CDM1), $ -0.0018\pm 0.0022$ (I$\Lambda$CDM2), $-0.1530^{+0.7400}_{-0.3900}$ (I$\Lambda$CDM3), and $-0.1100^{+0.1500}_{-0.1300}$ (I$\Lambda$CDM4). We can find that I$\Lambda$CDM2 gives the best constraints on the parameter $\beta$, followed by I$\Lambda$CDM4, while I$\Lambda$CDM1 and I$\Lambda$CDM3 give comparable constraint results. This is because in the early Universe, both $H$ and $\rho_{\rm c}$ have relatively high values, such that $\beta$ must be very small and can be constrained tightly. Therefore, in the I$\Lambda$CDM2 model, as a probe of the early Universe, CMB data can provide relatively stringent constraints on $\beta$. Furthermore, we investigate the impact of including the PR4 lensing data on parameter constraints. Our results show that in the $\Lambda$CDM model, CMB data (adding PR4 lensing data to the Planck 2018 data) results in $\sigma(H_0)=0.52~\rm km~s^{-1}~Mpc^{-1}$ and $\sigma(\Omega_{\rm m})=0.007$, which are 14.7\% and 15.2\% better than those of Planck 2018 data \citep{Wang:2021kxc}, respectively. However, in the IDE models, the addition of PR4 lensing data only slightly improves the parameter constraints. 

In Figure~\ref{fig2}, we show the constraint results of the CMB+DESI data in the $\beta$--$H_0$ (left panel) and $\beta$--$\Omega_{\rm m}$ (right panel) planes for the four IDE models. The constraint values of $\beta$ are $0.2900\pm 0.1810$ (I$\Lambda$CDM1), $ 0.0011\pm 0.0013$ (I$\Lambda$CDM2), $0.3600^{+0.3700}_{-0.2800}$ (I$\Lambda$CDM3), and $0.0750\pm 0.0520$ (I$\Lambda$CDM4). We find that the combinations including DESI data tend to favor slightly higher values of $H_0$ and lower values of ${\Omega_{\rm m}}$ compared to the central values derived from the CMB data. In this case, $\beta$ is positively correlated with $H_0$, as clearly seen in the left panel of Figure~\ref{fig2}, thus resulting in higher central values of $\beta$. Higher positive values of $\beta$ further support the scenario where CDM decays into dark energy. For instance, the I$\Lambda$CDM1 model with $\beta = 0.2900\pm 0.1810$ provides supporting evidence for an interaction, and the deviation of $\beta$ from 0 reaches $1.6\sigma$. Conversely, in the I$\Lambda$CDM2 model, the value of $\beta$ being close to zero indicates no interaction between dark energy and CDM.

In Figure~\ref{fig3}, we show the constraint results of the CMB+DESI+DESY5 data in the $\beta$--$H_0$ (left panel) and $\beta$--$\Omega_{\rm m}$ (right panel) planes for the four IDE models. The combination of low-redshift data and CMB data can effectively break the cosmological parameter degeneracies. For example, in the I$\Lambda$CDM2 model, CMB+DESI+DESY5 results in $\sigma(\beta)=0.0011$, $\sigma(H_0)=0.56~\rm km~s^{-1}~Mpc^{-1}$, and $\sigma(\Omega_{\rm m})=0.007$, which are 50\%, 62.6\%, and 67.1\% better than those of CMB, respectively. The $\beta$ values are $-0.1970\pm 0.0710$ (I$\Lambda$CDM1), $ 0.0003\pm 0.0011$ (I$\Lambda$CDM2), $-0.4000\pm 0.1300$ (I$\Lambda$CDM3), and $-0.0670\pm 0.0380$ (I$\Lambda$CDM4) when the CMB+DESI+DESY5 data are employed. {The deviations of $\beta$ from 0 for the I$\Lambda$CDM1, I$\Lambda$CDM2, I$\Lambda$CDM3, and I$\Lambda$CDM4 models reach significance levels of $2.7\sigma$, $0.3\sigma$, $3\sigma$, and $1.7\sigma$, respectively. In particular, we find that in the I$\Lambda$CDM2 model, $\beta$ is closer to 0 compared to the result obtained from CMB+DESI, which further demonstrates no interaction between dark energy and CDM. In other IDE models, it is indicated that there is an interaction where dark energy decays into CDM. Overall, these IDE models with $Q \propto \rho_{\rm de}$ are more inclined to support the presence of interaction compared to those with $Q\propto\rho_{\rm c}$. 

Under the current observational data, if an interaction exists, it involves the conversion of dark energy into dark matter. One might wonder whether this scenario could result in dark matter dominating over dark energy at some point in the future. Theoretically, IDE models may exhibit unphysical results in specific cases. For example, for $Q \propto \rho_{\rm de}$, if $\beta$ is positive, dark matter decays excessively into dark energy, causing the dark matter density to become negative at some future time (for more detailed discussions, see, e.g., \citealp{LopezHonorez:2010esq,He:2010im,Li:2013bya}). In our work, the value of $\beta$ we obtained is negative, which indicates that observations support the scenario where dark energy decays into dark matter. Therefore, we do not need to worry about the dark matter density becoming negative in the future. In fact, we also do not need to worry about the dark energy density becoming negative in the future. Here, we take the $Q=\beta H\rho_{\rm de}$ case as an example, and we have $\rho_{\mathrm{de}}=\rho_{\mathrm{de} 0} a^{-3(1+w)+\beta}$ for a constant $w$, where $\rho_{\mathrm{de} 0}$ is the current value of dark energy density; it can be seen that the density of dark energy will not become negative in the future. Moreover, we find that the current observational data only allow for a rather weak interaction, resulting in the evolution of energy densities in the IDE models being similar to that in the $\Lambda$CDM model. Therefore, in the future, the fraction of dark energy density will remain higher than that of dark matter, with dark energy continuing to dominate the evolution of the Universe.

In order to better understand the role played by DESI data in IDE models, we compare the theoretical distance predictions of $\Lambda$CDM and two typical IDE models (I$\Lambda$CDM1 and I$\Lambda$CDM2) against the observed cosmic distances in Figure~\ref{fig4}. Due to the relatively weak constraints provided by DESI alone, we combine DESI with CMB for our study. We analyze and compare the best-fit predictions from CMB+DESI for three different types of (rescaled) distances, including the angle-averaged distance ($D_{\mathrm{V}}$), the transverse comoving distance ($D_{\mathrm{M}}$), and the Hubble horizon ($D_{\mathrm{H}}$) derived from DESI BAO measurements. Foremost, we observe that the theoretical distance predictions of the $\Lambda$CDM and I$\Lambda$CDM2 models are largely consistent, indicating that the I$\Lambda$CDM2 model is largely compatible with the null hypothesis, i.e., the $\Lambda$CDM model. Conversely, the theoretical distance predictions of the $\Lambda$CDM and I$\Lambda$CDM1 models exhibit some deviations, indirectly suggesting the presence of interactions in I$\Lambda$CDM1. Moreover, we can see that the two data points deviating from the theoretical distance predictions of the $\Lambda$CDM and I$\Lambda$CDM2 models are the measurements of $D_{\mathrm{M}}/(r_{\mathrm{d}}z^{2/3})$ at $z = 0.71$ and $D_{\mathrm{H}}/(r_{\mathrm{d}}z^{-2/3})$ at $z = 0.51$. Interestingly, while the point $D_{\mathrm{H}}/(r_{\mathrm{d}}z^{-2/3})$ at $z = 0.51$ deviates from the theoretical predictions of the I$\Lambda$CDM1 model, I$\Lambda$CDM1 demonstrates greater success than the $\Lambda$CDM model in explaining $D_{\mathrm{M}}/(r_{\mathrm{d}}z^{2/3})$ at $z = 0.71$. Overall, apart from the $D_{\mathrm{M}}/(r_{\mathrm{d}}z^{2/3})$ and $D_{\mathrm{H}}/(r_{\mathrm{d}}z^{-2/3})$ at $z = 0.51$, all other BAO measurements align closely with the best-fit predictions of the I$\Lambda$CDM1 model.

Finally, we compare the IDE models based on their fittings to the current observational data using the Akaike information criterion (AIC; \citealp{Akaike:1974vps}), where $\mathrm{AIC} \equiv \chi_{\min }^{2}+2 k$ and $k$ represent the number of free parameters. In order to show the differences of AIC values between the $\Lambda$CDM and IDE models more clearly, we set the $\Delta {\rm AIC}$ value of the $\Lambda$CDM model to be zero, and list the values of $\Delta {\rm AIC}=\Delta \chi^2+2\Delta k$ in Table~\ref{tab2}, with $\Delta \chi^2=\chi^{2}_{\rm min, IDE}-\chi^{2}_{\rm min, \Lambda CDM}$ and $\Delta k=k_{\rm IDE}-k_{\rm \Lambda CDM}=1$. A model with a lower value of $\Delta {\rm AIC}$ is deemed to be more supported by the observational data.

When the CMB+DESI+DESY5 data are employed to constrain the IDE models, the $\Delta$AIC values are $-7.254$ (I$\Lambda$CDM1), $0.638$ (I$\Lambda$CDM2), $-10.628$ (I$\Lambda$CDM3), and $-2.718$ (I$\Lambda$CDM4). The I$\Lambda$CDM2 model, which has a slightly higher $\Delta {\rm AIC}$ than the $\Lambda$CDM model, indicates that its increased model complexity is not statistically supported when fitting the CMB+DESI+DESY5 data. Other IDE models have lower $\Delta {\rm AIC}$ values than the $\Lambda$CDM model, with the I$\Lambda$CDM4 model being only marginally lower. In general, the current observational data show a preference for IDE models where $Q\propto\rho_{\rm de}$ over those where $Q\propto\rho_{\rm c}$. Especially for the I$\Lambda$CDM3 model ($Q=\beta H_0 \rho_{\rm de}$), its $\Delta {\rm AIC}$ value is significantly lower than that of the $\Lambda$CDM model, and we can conclude that this model has more advantages than the $\Lambda$CDM model.

\section{Conclusion}\label{sec4}
In this work, our aim is to determine whether there is an interaction between dark energy and dark matter and to identify which IDE model is more supported by the current observational data. We use DESI, CMB, and DESY5 data to constrain the IDE models. Four phenomenological IDE models are considered: I$\Lambda$CDM1 ($Q=\beta H\rho_{\rm de}$), I$\Lambda$CDM2 ($Q=\beta H\rho_{\rm c}$), I$\Lambda$CDM3 ($Q=\beta H_0\rho_{\rm de}$), and I$\Lambda$CDM4 ($Q=\beta H_0\rho_{\rm c}$).

We find that using CMB data alone cannot provide strong constraints on IDE models due to the degeneracy among parameters. When DESI data are added, the central value of $\beta$ increases as the central value of $H_0$ increases, because $\beta$ is positively correlated with $H_0$. Specifically, the I$\Lambda$CDM1 model with $\beta = 0.2900\pm 0.1810$ provides supporting evidence for an interaction, and indicates a deviation from the standard $\Lambda$CDM model. Additionally, the best-fit predictions of the I$\Lambda$CDM1 model align closely with the majority of BAO measurements and demonstrate greater success than the $\Lambda$CDM model in explaining $D_{\mathrm{M}}/(r_{\mathrm{d}}z^{2/3})$ at $z = 0.71$. When combining DESI, CMB, and DESY5 data, the constraints on cosmological parameters can be significantly improved because this combination effectively breaks the degeneracies among these parameters. Furthermore, we find that $\beta$ is very close to zero, indicating no interaction between dark energy and CDM in the I$\Lambda$CDM2 model. Other IDE models suggest that there is an interaction where dark energy decays into CDM. For all IDE models, the level of deviations from $\Lambda$CDM always falls between $0.3\sigma$ and $3\sigma$. In particular, significant deviations from $\Lambda$CDM are observed for the I$\Lambda$CDM1 and I$\Lambda$CDM3 models. Generally, IDE models with $Q \propto \rho_{\rm de}$ support the existence of the interaction, while those with $Q \propto \rho_{\rm c}$ depend on the form of the proportionality coefficient $\Gamma$ to determine whether the interaction exists.

We use current observational data, including DESI, CMB, and DESY5 data, to calculate the AIC values for the four forms of IDE models. We find that the AIC values for the I$\Lambda$CDM1 and I$\Lambda$CDM3 models are significantly lower compared to $\Lambda$CDM and other IDE models, especially for the I$\Lambda$CDM3 model ($Q=\beta H_0 \rho_{\rm de}$) with $\Delta$AIC = $-10.628$. We can conclude that the current observational data show a preference for IDE models where $Q\propto\rho_{\rm de}$, and the I$\Lambda$CDM3 model has more advantages than the $\Lambda$CDM model in terms of fitting the observational data. In the coming years, the full DESI BAO dataset combined with CMB and SN datasets will contribute to a better understanding of the nature of dark energy. Therefore, pursuing this research is worthwhile, as future investigations may provide the necessary data to reassess this issue.

\section*{Acknowledgments}

We thank Yun-He Li, Yue Shao, Ji-Guo Zhang, and Bo Wang for helpful discussions. This work was supported by the National SKA Program of China (grants Nos. 2022SKA0110200 and 2022SKA0110203), the National Natural Science Foundation of China (grants Nos. 12473001, 11975072, 11875102, 11835009, and 12305068), the National 111 Project (grant No. B16009), and the China Scholarship Council.

\bibliography{DESI_IDE}
\end{document}